\documentclass[12pt]{article}
\usepackage{psfig,epsfig}
\title{Interaction between Physics and Cosmology}
\author{N. Panchapakesan\thanks{E--mail : panchu@bol.net.in} 
  	\\{\em Department of Physics and Astrophysics,} \\
	{\em University of Delhi, Delhi-110 007, India}
	}

\begin {document}
\maketitle
\begin {section}*{Abstract}
   Recent results indicate the presence of a cosmological constant (or related dark energy) in the universe. It has been conjectured recently that the interaction parameters of physical theories may be dependant on the size parameter of the universe, related to the cosmological constant. We investigate whether such effects will help in explaining baryogenesis in early universe. They do seem to succeed.
\end {section}

\begin {section}*{I. Introduction}

             Our knowledge of cosmology has improved remarkably in recent years. along with it has come many new or revived ideas and concepts. Recent observation of accelerated expansion of the universe requires the presence of a cosmological constant, introduced and withdrawn by Einstein long ago. The cosmological constant is related to the enegy of vacuum. Its small value is difficult to understand in particle physics.The cosmological constant leads to a de Sitter universe with a finite size; the larger the constant,the smaller the universe. Bjorken [1] has extended this idea of a relation between size of universe and the vacuum energy to other physical parameters which were earlier taken to be constant. During the period of inflationary expansion of the universe the vacuum energy is different and so is the size of the universe. We expect physical interaction parametrs to be also different.Inflationary expansions may occur during phase transitions in early universe.Baryogenesis in early universe is expected to occur during such phase transitions.If physical parameters change, as suggested by Bjorken, then this will affect baryogenesis. We shall consider this effect in this paper. 
\end {section}
\begin {section}*{II. Variaton of Standard Model Parameters}
   The size of the universe R and the value of cosmological constant $ \Lambda $ are related by the equation $\Lambda /3 = 1/R_\infty ^2 = 8\pi G \rho /3 = H_\infty^2 $
where the subscript refers to the values in the limit of very large times.
Bjorken [1] assumed that "All dimensionful parameters X of the standard model may vary with $ R_\infty$ but that to leading approximation they are straight lines in a log- log plot, i.e. they satisfy a simple renormalisation group equation". The equation satisfied is of the form:

  $ R^2 \frac {\partial X}{\partial R^2} = -1/2 \mu \frac {\partial X} { \partial \mu} =p_X X + .....$ .
  
where we have dropped the subscript on R.
\vskip 0.5 cm
      We consider the behaviour of  $ \Lambda_{QCD}$,  the cut of mass in QCD, v,the Vacuum expectation value of the Higgs field in electroweak theory and $ M_{Gut}$, the unification energy in grand unified theories. The parameters $ \Lambda_{QCD}, v, M_{Gut} $ all vary with R. This behaviour is shown in Fig.1. According to present ideas these parameters are all constant and never go up to $M_{Pl}$ as shown in the figure.
        
	As  $ 1/(\alpha _s (q^2)) = b_s ln (q^2 / (\lambda ^2))$

     and $ \Lambda_{QCD}^2 = (M_{Pl}^2 R^2)^{p_s} M_{Pl}^2$ 
     
     with $p_s \approx -1/3$, from Fig.1, it follows that 
    
     $ 1/(\alpha _s (q^2,R^2)) =b_s p_s ln (M_{Pl}^2 R^2) -b_s ln (M_{Pl}^2/(q^2)) $
These equations imply a renormalisation group equation for the coupling constant $\alpha_s$  of the form :
 \\ $ R^2 \frac{\partial} {\partial R^2} (1/(\alpha_s)) = b_s p_s + O (\alpha_s)$.\\

\vskip 0.5 cm
This implies a logarithmic increase for the coupling constant as R decreases.
When $ R \rightarrow \infty $ the coupling constant vanishes. In an infinite universe the standard model trivialises to a free field theory. All interacrions depend on the existence of a boundary to the universe. In Planck limit fields becomes strongly coupled.
The masses of quarks are given by $ m = g v $. So $m/v$ increases logrithmically as R decreases.
\end {section}
%\begin {figure}
%\vskip 5 truecm
%\special{psfile=jaipurfig.ps vscale= 70 hscale= 70}
%\epsfig {figure=Jaipurfig.eps, height=95mm}
%\caption {Dependance of physical parameters on size of universe}
%\label{fig:Dependance}
%\end {figure}

\begin {section}* {III. Baryon Assymmetry in the Universe}
There are many scenarios for baryogenesis. We shall confine ourselves to the one that occurs at the time of electroweak phase transition [2][3]. This scenario uses only parameters of the standard model known from experiments.In this model [2] the baryon aymmetry of the universe (BAU) is given by

   $BAU = J[(m_t)^2 - (m_u)^2)((m_t)^2-(m_c)^2)((m_c)^2-(m_u)^2)((m_b)^2-(m_s)^2)
           ((m_b)^2-(m_d)^2) ((m_s)^2-(m_d)^2) / (T^{12}] $.
     
     $ J= \sin(\theta_{12}) \sin(\theta_{13}) \sin(\theta_{23})\sin(\delta_{CP})$.
        
Here J is the CP violating factor determined from decays of K and now B mesons.
The angles are mixing angles of CKM matrix for quark mixing [2].

    This BAU has a value of $10^{-21}$ while the required value based on nucleosynthesis in early universe is of the order of $10^{-10}$
    
    The new point we are trying to make is that the value of R at the time of phase transition is different due to supercooling and inflationary expansion. The value of R is  ~ 1 cm at the e.w. phase transition, while the value of R for our universe is $10^{28} cm $. The physical parameters of the standard model change with R as discussed in the previous section. This will give an additional factor of $ (log (R/R_{ew})^{12}$ in BAU. The value of this factor is $10^{16}$. So the BAU is about $10^{-5}$ which is a much more reasonable value than the earlier one, as detailed analysis will bring it down by a few orders.
    
      There is thus hope that baryogenesis scenario at the time of elactroweak phase transition may explain the value of BAU that is required [4].    
      
\end {section}
\begin {section}*{IV Discussion}
      As mentioned earlier there are many scenarios for baryogenesis [3].We have considered the one closest to known physics in the laboratory. This scenario has two major difficulties. The first is the requirement of a  small mass for the Higgs particle to ensure first order phase transition. This seems already inconsistent with the observations. To study this difficulty in the present framework, the constraints on the mass of the Higgs have to be reworked taking the variation of the parameters at the time of phase transition. This requires detailed calculations along the lines done in our earlier work[2]. The second difficulty is the very small value of BAU that is predicted. We have shown above that the second difficulty seems to be overcome if parameters depend on the size of the universe. 

       The simplest extension of our scenario involves minimal super- symmetry with an extra Higgs particle. This has to await signals of super- symmetry in the laboratory. The other scenarios assume  fields whose parameters ( even their very existence) is unobserved and unconfirmed in thelaboratory. The scenarios based on string theories and brane models are again in a different realm.
\end {section}
\begin {section}*{V References} 

 1. J.D.Bjorken,  Phys. Rev. D 67,043508-18 (2003).

2. A.S.Majumdar,S.K.Sethi,S.Mahajan, A.Mukherjee, N.Panchapakesan,and R.P.Saxena, Mod. Phys. Lett., A9,459 (1994).
 G.R.Farrar,and M.E.Shaposhnikov  Phys.Rev. Lett.,70,2833 (1993), 71,210(E) (1993), Phys.Rev. D50, 774 (1994).

3.M.Trodden, Rev. Mod. Phys.       (1999), A.Riotto and M.Trodden,hep-phys/9901362,
  M.Quiros, hep-phys/9901312.

4. For a similar approach but with a different motivation, see  M.Berkooz, Y.Nir, and T. Volansky, Phys. Rev. Lett. 93,051301-4 (2004).
 
\end {section}
\begin {figure}
\vskip 5 truecm
\includegraphics{jaipurfig.ps}
\caption {Dependance of physical parameters on size of universe}
\label{fig:Dependance}
\end {figure}
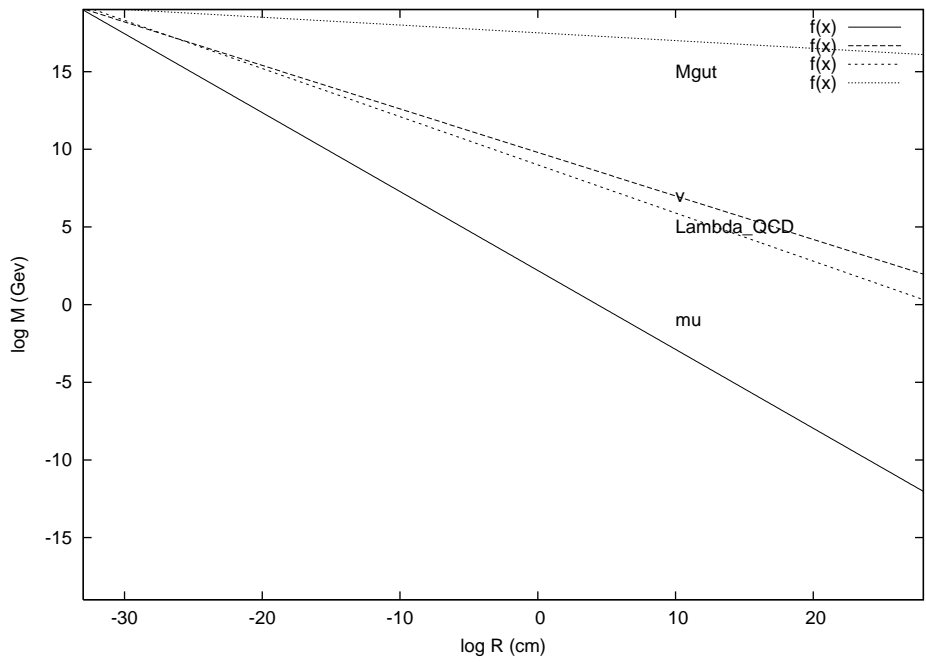
\end{document}